\documentclass[twocolumn,showpacs,preprintnumbers,amsmath,amssymb,superscriptaddress]{revtex4}

\usepackage[colorlinks=true]{hyperref}
\usepackage{graphicx}
\usepackage{units}
\usepackage{bm}

\begin{document}


\title{Contacting individual Fe(110) dots in a single electron-beam lithography step}



\author{F. Cheynis}
\email[]{cheynis@cinam.univ-mrs.fr}
\altaffiliation[Present adress: ]{Centre Interdiscipliaire de Nanoscience de Marseille, CNRS - UPR 3118
Campus de Luminy Case 913, 13288 Marseille Cedex 9, France}
\affiliation{Institut N\'{E}EL, CNRS-UJF, BP 166, 38042 Grenoble Cedex 9, France}
\affiliation{Grenoble-INP, 46 avenue F\'{e}lix Viallet, 38031 Grenoble Cedex 1, France}

\author{H. Haas}
\affiliation{Institut N\'{E}EL, CNRS-UJF, BP 166, 38042 Grenoble Cedex 9, France}

\author{T. Fournier}
\affiliation{Institut N\'{E}EL, CNRS-UJF, BP 166, 38042 Grenoble Cedex 9, France}

\author{L. Ranno}
\affiliation{Institut N\'{E}EL, CNRS-UJF, BP 166, 38042 Grenoble Cedex 9, France}

\author{W. Wernsdorfer}
\affiliation{Institut N\'{E}EL, CNRS-UJF, BP 166, 38042 Grenoble Cedex 9, France}

\author{O. Fruchart}
\affiliation{Institut N\'{E}EL, CNRS-UJF, BP 166, 38042 Grenoble Cedex 9, France}

\author{J.-C. Toussaint}
\affiliation{Institut N\'{E}EL, CNRS-UJF, BP 166, 38042 Grenoble Cedex 9, France}
\affiliation{Grenoble-INP, 46 avenue F\'{e}lix Viallet, 38031 Grenoble Cedex 1, France}


\date{\today}

\begin{abstract}
We report on a new approach, entirely based on electron-beam lithography technique, to contact electrically, in a four-probe scheme, single nanostructures obtained by self-assembly. In our procedure, nanostructures of interest are localised and contacted in the same fabrication step. This technique has been developed to study the field-induced reversal of an internal component of an \emph{asymmetric Bloch} domain wall observed in elongated structures such as Fe(110) dots. We have focused on the control, using an external magnetic field, of the magnetisation orientation within \emph{N\'{e}el caps} that terminate the domain wall at both interfaces. Preliminary magneto-transport measurements are discussed demonstrating that single Fe(110) dots have been contacted.
\end{abstract}


\maketitle

Contacting individual nanostructures to probe their intrinsic transport properties has been a major technological challenge within the last few years. Recently, nanowires have been widely studied for their quantum properties\cite{bib-CLE06,bib-BEZ00,bib-LUC07} and also as promising candidates for next generation nanoelectronic transistors and devices\cite{bib-HUA01,bib-THE06}. Most of these works have opted for a top-down approach mostly based on electron-beam (e-beam) lithography to contact electrically randomly distributed nanostructures. This route requires two technological steps that consist first in localising the nanostructures to be electrically probed and then contacts are made to individual wires using a lift-off technique. The first step is clearly time-consuming as AFM or SEM pictures are needed. Another approach has been developed for nanowires embedded in a membrane and obtained by electrodeposition\cite{bib-MIC03}. Membrane
holes are partially plugged using a physically-deposited top Au layer. The growth of nanowires is stopped abruptly when the first structure emerges from the membrane, leaving the other wires unconnected. The growth termination is detected by the sharp change in the deposition current. This simple low-cost technique is however restricted to a very specific system. Alternative routes have been developed using local probe microscopy techniques such as AFM\cite{bib-BOU03} or STM\cite{bib-IAC08}. The former makes it possible to indent accurately in depth and lateral position a photoresist layer. The indented void then serves as a mask to create top contact. The latter avoids any technological process as it defines \emph{per se} a tunnel junction between the tip and the probed nanostructure. These local probe techniques are nevertheless limited by their low throughput.

In this letter, we report on a new technique, entirely based on e-beam lithography, to carry out four-probe magneto-resistance measurements of self-assembled Fe(110) dots. We opted for a simple approach that can be applied as a general procedure to contact other self-assembled nanostructures randomly distributed over the sample surface. Our innovative approach makes it possible to circumvent the technical difficulties which consist in localising precisely self-assembled structures and in contacting selected structures. In our method, we addressed these two crucial issues in a single e-beam lithography step. Preliminary magneto-resistance (MR) measurements of individual Fe(110) dots are then presented as a proof of the reliability of our approach.\\


\begin{figure}[h]
		 \includegraphics[width=70mm]{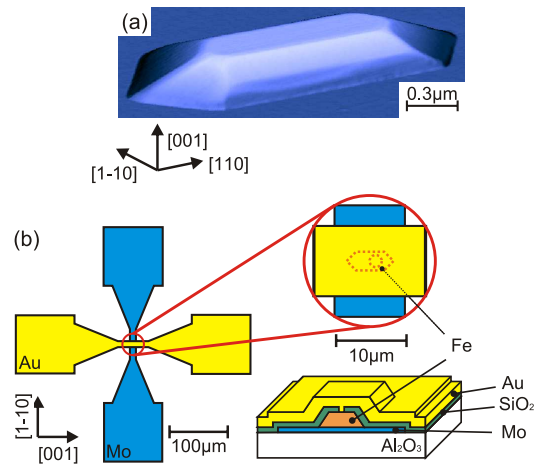}
		 \caption{\label{fig:Fig1} (a) AFM image of a typical Fe(110) dot obtained by PLD (Ref. \cite{bib-CHE08}). (b) Schematic illustration of the four-probe design with the bottom (Mo) and top (Au) electrodes. These electrodes are electrically isolated by a SiO$_{2}$ layer. The dielectric layer has to be etched on top of an individual selected Fe(110) dot in the central area of the pattern.}
	\end{figure}

Fe(110) dots are obtained by Pulsed-Laser Deposition (PLD) under Ultra-High Vacuum conditions. These structures grow epitaxially on a Mo(110) buffer layer previously deposited on a sapphire (11$\bar{2}$0) substrate. Self-assembly results from the so-called Stranski-Krastanov growth mode, occurring for a deposition temperature in the range $\sim$\unit[600-850]{K} [\ref{fig:Fig1}(a)]\cite{bib-FRU06}. For our study, the Fe dots have been deposited at \unit[$\sim$800]{K} on a \unit[50]{nm} Mo(110) buffer layer and covered by a \unit[$\sim$0.7]{nm} Mo layer and a \unit[5]{nm} Au layer to prevent from oxydation. 

The sample has been processed using a Field Emission Gun Scanning Electron Microscope (FEG-SEM) \emph{LEO 1530} remote-controlled by the \emph{Elphy plus} system (\emph{Raith GmbH}) for e-beam lithography (typical e-beam voltage of \unit[20]{kV}). A schematic view of the pattern is presented in \ref{fig:Fig1}(b). The first step of our process is based on a routine procedure for the fabrication of the bottom electrodes in the Mo(110) buffer layer by Ar Ion Beam Etching using \emph{Shipley UVN2$^\circledR$} chemically amplified negative resist as a mask. The \unit[230]{nm} resist mask is thick enough to protect \unit[$\sim$140]{nm}-high Fe(110) dots upon the etching process of the \unit[50]{nm} Mo(110) buffer layer. The UVN2 layer is removed using \emph{EKC$^{\circledR}$LE$^{\circledR}$} solution and a \unit[55]{nm} SiO$_{2}$ layer is then deposited on the sample surface using RF Magnetron sputtering technique. 


\begin{figure}[h]
		 \includegraphics[width=85mm]{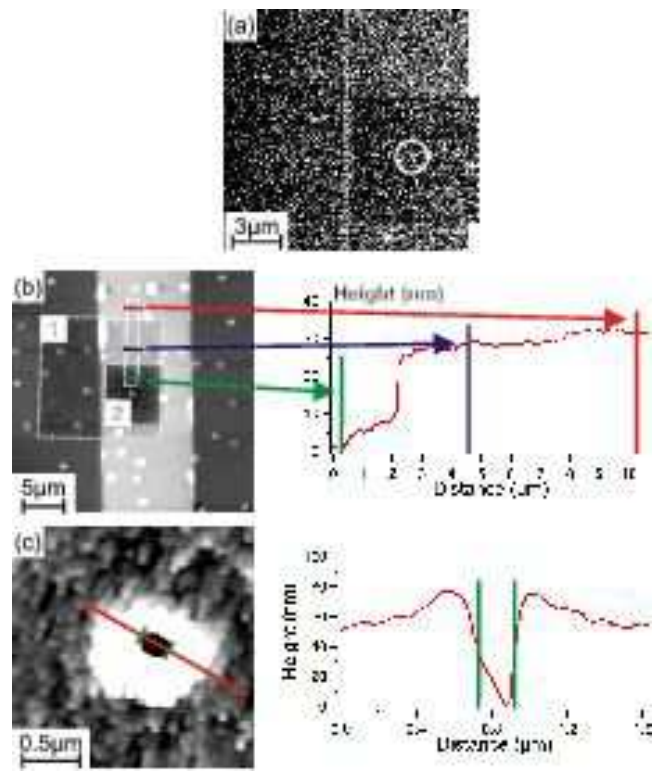}
		 \caption{\label{fig:Fig2} Optimised imaging conditions of the central area of our pattern using a standard \unit[20]{kV} e-beam voltage. (a) SEM images of the sample surface covered by a PMMA resist layer and a \unit[10]{nm} Al layer. These images have been used to shift the e-beam writefields on top of a selected Fe(110) dot (white circle). The dashed white line is a guide-to-the-eye for the limit of the bottom contacts. (b-c) Contact AFM images of the central area of the \emph{test} sample obtained after imaging, selection, exposure and development steps. Image (b) makes it possible to quantify the effect of imaging the PMMA$\backslash$Al surface on the resist thickness. Image (c) highlights the position accuracy of the exposure step and the dimensions of the exposed area.}
	\end{figure}

The next step of the process consists in patterning the SiO$_{2}$ layer on top of selected Fe(110) dots ($\phi=\unit[100]{nm}$ and $\phi=\unit[200]{nm}$ holes) and on top of the bottom contact pads using a \emph{Microchem PMMA} positive resist layer as a mask. The key aspect of our process has been to develop an exposure procedure to localise and contact individual Fe(110) dots in a single e-beam step. This approach is based on the imaging of the central area of the pattern covered with a PMMA resist layer. A single Fe(110) dot per pattern is then selected and exposed with a view to creating a hole in the resist layer in the very same process step. The technical difficulty of this step has been to optimise the quality of the image used to determine the structure to be connected without exposing the electron sensitive PMMA resist in a too significant manner. The procedure has been optimised using a standard \unit[180]{nm} PMMA \unit[3]{\%} layer spin-coated on a so-called \emph{test} sample exhibiting \unit[$\sim$120]{nm} Fe(110) dots [\ref{fig:Fig2}]. In our process, two images are required to shift e-beam writefields with a sufficient accuracy ($<$\unit[150]{nm}) on top of the selected structure. The sample surface is imaged under the same e-beam voltage conditions as that used for exposure steps (\unit[20]{kV}). The key parameters of the imaging conditions are the size (in $\mu$m) and the resolution (in px) of each image. These parameters determine, knowing the beam current and the dwell-time per pixel, an equivalent exposure dose (EED). SEM imaging of Fe(110) dots is possible because the PMMA surface is not perfectly flat on top of Fe(110) dots. A \unit[10]{nm} Al layer deposited on top of the resist layer is also used to allow imaging of the surface of the sample covered with an insolating PMMA layer. Typically, \unit[30-40]{nm}-high prominences have been observed using AFM in contact mode. The thickness of the PMMA layer on top of Fe(110) dots is thus $\sim$\unit[80]{nm}.

	The effect of SEM image acquisition step on the final PMMA resist thickness and the accuracy of the e-beam writefields' shifts have been quantified by contact AFM images [\ref{fig:Fig2}(b-c)]. It turns out that the first \unit[15]{$\mu$m}$\times$\unit[15]{$\mu$m} SEM picture (\unit[256]{px}$\times$\unit[256]{px}), labelled \textbf{1} in \ref{fig:Fig2}(b), induces a loss of only \unit[3-4]{nm} of the resist layer while in the overlapping area with the second SEM image, labeled \textbf{2} in \ref{fig:Fig2}(b), \unit[$\sim$30]{nm} of PMMA have been removed. The remaining thickness of the PMMA layer on top of each exposed Fe(110) dot is thus \unit[50]{nm}, which turned out to be sufficient to protect the rest of the exposed Fe(110) dot as detailed below. The EED of each image is \unit[$\sim$27]{$\mu$C/cm$^{2}$} so that the total EED of \unit[$\sim$54]{$\mu$C/cm$^{2}$} remains significantly below typical exposure dose values of \unit[450-600]{$\mu$C/cm$^{2}$} used for PMMA resist layers. \ref{fig:Fig2}(c) highlights the fact that an exposure position accuracy better than \unit[150]{nm} is obtained. From this image, the thickness of resist layer on top of a Fe(110) dot can be estimated to \unit[$\sim$70]{nm}. This thickness is however underestimated owing to the finite AFM tip radius (typically \unit[10]{nm}) and the full tip cone angle (typically \unit[40]{$^{\circ}$}). This is consistent with the estimate of a resist layer thickness of \unit[80]{nm} meaning that the resist has been completely exposed on top of the selected Fe(110) dots. It also turns out that the FWHM of the exposed area is of the order of \unit[205]{nm} which is good agreement with a nominal diameter of \unit[200]{nm}.
	
	For the final sample exhibiting Fe(110) dots of typical height \unit[$\sim$140]{nm}, we have used a thicker resist layer (PMMA \unit[4]{\%}, \unit[240]{nm}). Although the PMMA \unit[4]{\%} layer has been observed to be more sensitive to e-beam exposure, similar protection conditions of the \unit[$\sim$140]{nm} Fe(110) dots have been obtained.  
	
\begin{figure}[h]
		 \includegraphics[width=75mm]{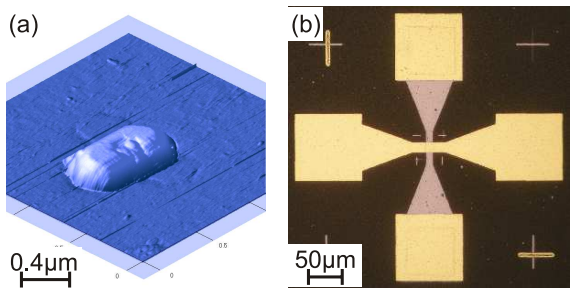}
		 \caption{\label{fig:Fig3} (a) Contact mode AFM image of a Fe(110) dot after an extended etching process. (b) Optical image of the final pattern highlighting bottom (grey) and top (yellow) contacts.}
	\end{figure}

	Once developed, the PMMA resist layer serves as a mask for a Reactive Ion Etching of the SiO$_{2}$ layer using a CHF$_{3}$ RF plasma ($P_{CHF_{3}}=\unit[2.10^{-2}]{Torr}$, $P_{RF}=\unit[50]{W}$). The etching has been monitored using laser reflectometry performed on a Si$\backslash$SiO$_{2}$ reference layer deposited together with the covering layer of the Fe(110) dots. For the test sample exhibiting \unit[$\sim$120]{nm} Fe(110) dots, the etching process has been intentionally extended. \ref{fig:Fig3}(a) clearly demonstrates that only the area corresponding to the exposure pattern is affected by the etching process, leaving the area imaged using SEM unaffected.
	
	The last step of the process is a standard lift-off process of a Ti$\backslash$Au (\unit[10]{nm}$\backslash$\unit[100]{nm}) layer using an automatic realignment procedure and a PMMA positive resist [\ref{fig:Fig3}(b)]. The structured Ti$\backslash$Au layer finally defines the top electrodes and connects a single Fe(110) dot per pattern and the bottom electrodes.\\
	
Fe(110) dots have been contacted to observe in real-time the reversal of an internal component of a magnetic domain wall and demonstrate experimentally its associated hysteresis. Indeed, we have recently demonstrated that the magnetisation orientation within the two N\'{e}el caps (NC's) that terminate an asymmetric Bloch wall at both interfaces in elongated structures (in our case self-assembled Fe(110) dots) can be controlled using an external magnetic field\cite{bib-CHE08, bib-CHE08b}. The signal used to probe the NC reversal is the \emph{Anisotropic Magneto-Resistance} (AMR) which yields to two different resistance states for a transverse (\emph{i.e.} along the width of the Fe(110) dots) current component.

\begin{figure}[h]
		 \includegraphics[width=85mm]{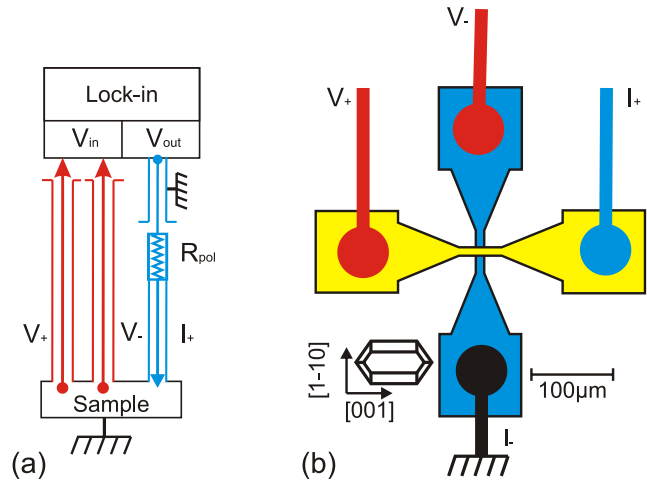}
		 \caption{\label{fig:Fig4} Schematic illustrations of the magneto-transport setup. (a) A \emph{Stanford Research Systems SR830} lock-in device and a high-impedance polarisation charge (\unit[1]{k$\Omega$}) are used to feed an AC current of constant amplitude through a single Fe(110) dot. The four-probe sample voltage is measured at the lock-in frequency. (b) Four-probe contacts of a pattern.}
	\end{figure}

Magneto-transport measurements have been carried out in a variable temperature (\unit[2.2]{K}-\unit[300]{K}) pumped He cryostat. An in-plane magnetic field up to \unit[7]{T} is provided by a superconducting split-coil magnet. The sample holder can be rotated continuously in-the-plane of the sample. In our measurements, we have used two applied field configurations with respect to Fe(110) dots' orientation : along the [001] (called \emph{longitudinal} hereafter) direction and along the [1$\bar{1}$0] (called \emph{transverse} hereafter) direction. Typical field scans in the range $\pm\unit[1.2]{T}$ have been recorded in \unit[10]{min}. Low-noise measurements are required as, according to our RT micromagnetic simulations\cite{bib-CHE08b}, the transverse magnetisation averaged over the 3D structure is expected to decrease by \unit[4]{\%} upon NC reversal. Furthermore in Ref.\cite{bib-TON93}, the authors determine the RT AMR signal for an epitaxial Fe$\left\{110\right\}$ layer with a current applied along the [110] direction (\emph{i.e.} an equivalent direction to our \emph{transverse} configuration) of the order of \unit[0.3]{\%}. This yields to a RT-MR signal of the order of 10$^{-4}$, which, in the case of an impedance of $\sim\unit[1]{\Omega}$, is a severe experimental difficulty. Our low-noise/low-impedance setup is based on a \emph{Stanford Research Systems SR830} lock-in device and a high-impedance polarisation charge (\unit[1]{k$\Omega$}) as a current source (\ref{fig:Fig4}). The four-probe sample voltage is then detected at the lock-in frequency (\unit[1013]{Hz}). In our setup, peak-to-peak relative noise amplitude has been reduced down to $\sim$7$\times$10$^{-5}$ at \unit[1013]{Hz} which is close to the lock-in limit measured at  $\sim$5$\times$10$^{-5}$ in this frequency range.

\begin{figure}[h]
		 \includegraphics[width=85mm]{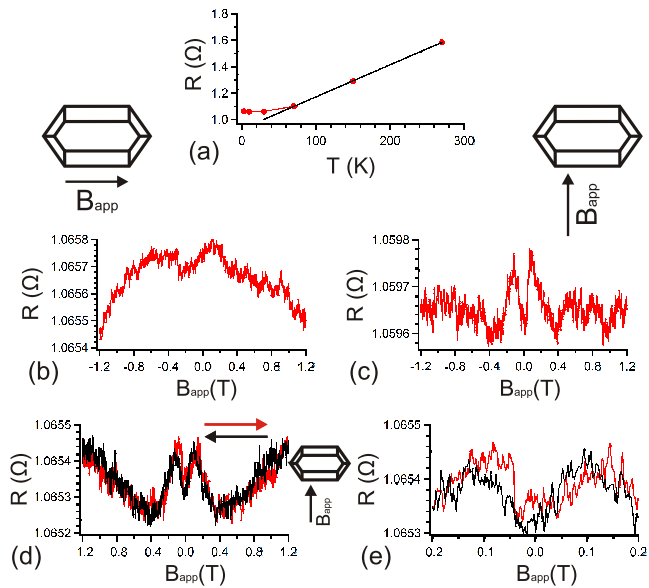}
		 \caption{\label{fig:Fig5} (a) Zero-field resistance of a single Fe(110) dot as a function of sample temperature. (b-d) Magneto-resistance for (b) longitudinal and (c) transverse applied field directions. (d) Complete field-scan (back and forth between \unit[-1.2]{T} and \unit[+1.2]{T}) magneto-resistance under a transverse magnetic field. (e) Zoom-in of (d) in the \unit[$\pm$0.2]{T} range. For all measurements, the typical injection current is $\unit[\simeq1]{mA}$.}
	\end{figure}

The zero-field resistance of a pattern is plotted in Fig. \ref{fig:Fig5} as a function of sample temperature. It clearly exhibits a linear variation with $T$ from \unit[$\sim$1.6]{$\Omega$} at room temperature (RT) down to \unit[$\sim$1.1]{$\Omega$} at \unit[50-70]{K} where it reaches a residual value. This demonstrates that a metallic contact has been established between the top and bottom electrodes. This metallic contact cannot be attributed to a leakage through the dielectric layer which has been observed to yield to typically higher contact resistances (\unit[$\sim$50]{$\Omega$}) for defective samples with no hole, obtained during the lithography optimisation procedure. To confirm this, the expected device resistances have been derived by 3D finite-element electrostatic simulations using \emph{COMSOL Multiphysics} software. In a purely diffusive model, RT bulk resistivities of the different materials have been used. It turns out that resistances of \unit[0.3]{$\Omega$} and \unit[0.2]{$\Omega$} are expected respectively for the Au contact cylinder on top of the connected Fe(110) dot and for the Fe(110) dot itself. Despite working in a four-probe configuration, a resistance of \unit[0.5]{$\Omega$} arising from the Mo layer is predicted. A lead resistance of the same order of magnitude is expected in the Au layer yielding to a total extra resistance of \unit[$\sim$1.3]{$\Omega$}. A lead resistance in low-impedance structures is known as \emph{current-crowding}\cite{bib-CHE02}. Using Eq. (7) in Ref.\cite{bib-CHE02}, we find an additional lead resistance of \unit[$\sim$1.4]{$\Omega$}. In both cases, the predicted four-probe resistance is in quantitative agreement with experimental value. This is consistent with the contacting of a single Fe(110) dot.

Typical low-temperature MR measurements obtained for longitudinal and transverse fields are shown in Fig. \ref{fig:Fig5}(b-c). Fig. \ref{fig:Fig5}(b) clearly exhibits a Lorentz-like MR (\emph{i.e. $\sim B_{\mathrm{app}}^{2}$}) despite a counter-intuitive negative component. The origin of this negative MR remains an open question. The MR behaviour of the temperature sensor may play a role. Peaks in the MR response are observed only for transverse applied fields [Fig. \ref{fig:Fig5}(c)]. More particularly, the minima measured at \unit[$\sim\pm$0.4]{T} may be ascribed to the transverse saturation of the Fe(110) dots in a quantitative agreement with hysteresis loops of an assembly of Fe(110) dots\cite{bib-FRU03c}. The two maxima observed at \unit[$\sim\pm$0.12]{T} are thus likely to be related to the NC's reversal, in qualitative agreement with micromagnetic simulations\cite{bib-CHE08b}. From Fig. \ref{fig:Fig5}(d-e), no clear evidence of the existence of an hysteresis associated with NC reversal can be drawn. An improvement would be to reduce the current crowding effect by etching trenches in electrodes around the selected nanostructure as suggested in Ref.\cite{bib-GIJ93}. Superconducting electrodes constitutes also a promising alternative route as lead resistances would be suppressed. In this perspective, Nb appears as a high potential candidate as Nb(110) buffer layers have been already obtained in our deposition chamber.\\ 

		In conclusion, we have developed a new approach to contact self-assembled nanostructures, by nature randomly distributed on a surface. This procedure make it possible to locate and expose individual structures in a single e-beam lithography step. It can be used as a general route to connect structures randomly distributed over the sample surface. This technique has been developed to study the field-induced reversal of an internal degree of freedom of an asymmetric Bloch wall (\emph{i.e.} the orientation of NC's) in individual Fe(110) dots. Preliminary MR results obtained confirm that individual Fe(110) dots have been connected  and that magnetic processes can be monitored through magnetotransport measurements based on the contacting process. 

\begin{acknowledgments}
We are grateful to P. David and V. Santonacci (sample elaboration), T. Crozes (nanofabrication), D. Dufeu and D. Maillard (magneto-transport), and D. Lucot for technical support and/or fruitful discussions.
\end{acknowledgments}


\end{document}